\documentclass[10pt]{iopart}
\usepackage{iopams}
\usepackage{epsfig}
\begin{document}
\title{Polarization states
of gravitational waves with a massive graviton}
\author[W L S de Paula {\it et al}]{Wayne L.S. de Paula,
Oswaldo D. Miranda and Rubens M. Marinho}
\address{Instituto Tecnol\'ogico de Aeron\'autica, Departamento de
F\'\i sica, Pra\c ca Marechal Eduardo Gomes 50, S\~ao Jos\'e dos
Campos, 12228-900 SP-Brazil}

\begin{abstract}
Using the Newman$-$Penrose formalism, we obtain the explicit
expressions for the polarization modes of weak, plane
gravitational waves with a massive graviton. Our analysis is
restricted for a specific bimetric theory  whose term of mass, for
the graviton, appears as an effective extra contribution to the
stress-energy tensor. We obtain for such kind of theory that the
extra states of polarization have amplitude several orders of
magnitude smaller than the polarizations purely general relativity
(GR), $h_{+}$ and $h_\times$, in the VIRGO$-$LIGO frequency band.
This result appears using the best limit to the graviton mass
inferred from solar system observations and if we consider that
all the components of the metric perturbation have the same
amplitude $h$. However, if we consider low frequency gravitational
waves (e.g., $f_{\rm GW}\,{\sim}\,10^{-7}{\rm Hz}$), the extra
polarization states produce similar Newman$-$Penrose amplitudes
that the polarization states purely GR. This particular
characteristic of the bimetric theory studied here could be used,
for example, to directly impose limits on the mass of the graviton
from future experiments that study the cosmic microwave background
(CMB).
\end{abstract}

\submitto{\CQG}


%\maketitle

\section{Introduction}

General Relativity (GR) assumes that gravitational forces are
propagated by a massless graviton. However, the present
experimental limits on the mass of the graviton are only based on
the behaviour of static gravitational fields as, for example, the
Newtonian planetary motion in the solar system. If gravity is
described by a massive graviton, the Newtonian potential would
have Yukawa modifications of the form

\begin{equation}
V(r)={GM \over r} \exp{(-r/\lambda_{\rm g})},
\end{equation}

\noindent where $M$ and $\lambda_{\rm g}$ are the mass of the
source and the Compton wavelength of the graviton, respectively.

The best bound on the graviton mass from planetary motion surveys
is obtained by using Kepler's third law to compare the orbits of
earth and mars, yielding $m_{\rm g} < 4.4 \times 10^{-22} {\rm
eV}$ \cite{larsonhiscock}. Another bound on the graviton mass can
be established by considering the motions of galaxies in clusters
of galaxies, yielding $m_{\rm g} < 2 \times 10^{-29} {\rm eV}$.
This second bound is less robust than solar system estimates due
to the uncertainty about the matter distribution of the universe
on large scales \cite{larsonhiscock, will98, will01}.

A graviton with nonzero mass would produce several effects in the
dynamical regime as, for example, extra degrees of polarization
for the generation of gravitational waves and velocities of
propagation dependent on the frequency of the waves.

Based on these characteristics \cite{will98} has suggested that
the mass of the graviton could be bounded using gravitational wave
observations. As binary systems evolve, they will slowly spiral
together due to the emission of gravitational radiation. Over the
course of time, the frequency of the binary orbit rises, ramping
up rapidly in the late stages of the evolution, just prior to
coalescence. Laser interferometer gravitational wave detectors
should be able to track the binary system's evolution, obtaining
the detailed time-dependent waveform using the matched filtering
techniques required for data analysis in these detectors.

At least in principle, another possibility to identify the effects
produced by massive gravitons consists in studying the excited
vibrational eigenmodes of spherical gravitational wave detectors
to identify the field content of a specific gravitational theory
by the observed features of the waves \cite{Bianchi}.

It is worth stressing that gravitational wave detectors, either
interferometers or resonant (cryogenic bars and spheres) are
members of a network that will permit the reduction of spurious
signals and an experimental determination of the false alarm rate.
In addition, three or more detectors ensure the complete
reconstruction of a gravitational wave event, including the
determination of its velocity of propagation and, the
identification of the Riemann tensor signatures.

From the theoretical point of view,  massive gravitons have a
number of strange properties produced at the level of the field
equations. If the mass term has a specific Fierz$-$Pauli structure
\cite{Nieuwenhuizen,fierzpauli}, the propagator around flat space
suffers the van Dam$-$Veltman$-$Zakharov (vDVZ) discontinuity
\cite{vandam,zakharov,boulwaredeser}.

On the other hand, the effects of Fierz$-$Pauli mass terms were
considered only around flat space backgrounds. Thus, the
discontinuity that was found could be a peculiarity of the flat
background and some of the difficulties could be evaded by
considering a background with curvature \cite{dkp02}.

In fact, it was found that in constant curvature backgrounds (AdS
spaces), the extra polarizations of the massive gravitons have a
coupling $\sim  m_{\rm g}/H$ where $H$ is the Hubble constant of
the AdS space \cite{higuchi,porrati}. In this case, the
predictions of the massive theory were the same as of the massless
theory when $m_{\rm g} \ll H$.

Nevertheless, it is important to stress that GR has an excellent
agreement in the prediction of the decrease of the orbital period
($\tau$) of the binary pulsar PSR 1913+16 at least in the weak
field limit. The GR predicts $\tau=-2.4\times 10^{-12}$ and the
observed value is $\tau=-(2.40243\pm 0.00005)\times 10^{-12}$
\cite{will01}. Thus, an alternative theory of gravity in the limit
$m_{\rm g}\rightarrow 0$ should obtain the same results of GR.
Besides, the Newtonian limit has to be valid.

In fact, theories of gravitation can be divided into two different
groups: metric and non-metric theories \cite{will93}. Basically, a
theory of gravitation is said to be metric if it obeys the
principle of equivalence. That is, the action of gravitation on
the matter is due exclusively to the metric tensor. GR and
Brans$-$Dicke theories are examples of metric theories of
gravitation.

Thus, one of the ways to obtain a massive graviton theory,
preserving the equivalence principle, is to add a prior geometry.
This possibility was recently proposed by \cite{matt98}. In his
theory, it is possible to recover GR when $m_{\rm g}\rightarrow 0$
without the problem of the vDVZ discontinuity because, the
linearized mass term of the theory is not a Pauli$-$Fierz term.

In the present study, we analyse the polarization states of the
gravitational waves, in the theory developed by \cite{matt98},
using the formalism developed by Newman$-$Penrose
\cite{newmannpenrose}. Then, we follow the method proposed by
\cite{eardleylightmanlee,lightmanlee} to determine the number of
polarization modes of the gravitational waves. The final result
consists in the explicit expressions for the six independent
`eletric' components of the Riemann tensor.

In section \ref{sec2}, we present the field equations for the
massive graviton using the action proposed by \cite{matt98}. In
section \ref{sec3}, we obtain the general solution for weak
gravitational waves with a massive graviton. In section
\ref{sec4}, we determine the explicit expressions for the
polarization modes of the gravitational waves using the
Newman$-$Penrose formalism. In section \ref{sec5} we present our
conclusions and discuss some aspects on how it could be possible
to impose limits on the mass of the graviton from observations of
the cosmic microwave background.

\section{Field equations for massive graviton}
\label{sec2}

We wish to examine an extension of linearized GR which includes a
massive graviton in the field equations. As exposed above, our
main idea is recover GR when $m_{\rm g}\rightarrow 0$ so, the
linearized mass term is not the Pauli$-$Fierz term. That is an
essential condition to have a well-behaved classical limit as the
graviton mass goes to zero. According to \cite{matt98} a way to
get this is to add the massive graviton term in the action as an
isolated term. That is

\begin{equation}
I=\frac{c^{4}}{16\pi G}\frac{I_{\rm G}}{c}+\frac{I_{\rm
F}}{c}+\frac{I_{\rm mass}}{c}. \label{action}
\end{equation}

The last term on the right-hand side of (\ref{action}) contributes
to the equations of motion with an effective stress tensor given
by

\begin{equation} T^{\mu\nu}_{\rm mass}={2\over \sqrt{-g}}{\delta\over \delta
g_{\mu\nu}}I_{\rm mass},
\end{equation}

\noindent where

\begin{eqnarray}
T^{\mu\nu}_{\rm mass} &=& -\frac{m_{\rm g}^{2}c^{6}}{8\pi
G\hbar^{2}}\bigg\{(g_{0}^{-1})_{\mu\sigma}\bigg[(g-g_{0})^{\sigma\rho}
\nonumber\\ &&
-\frac{1}{2}(g_{0})^{\sigma\rho}(g_{0}^{-1})_{\alpha\beta}
(g-g_{0})^{\alpha\beta}\bigg](g_{0}^{-1})_{\rho\nu}\bigg\},
\label{massa}
\end{eqnarray}

\noindent where $(g_{0})_{\alpha\beta}$ is the non-dynamical
background metric, $g_{\alpha\beta}$ is the physical (dynamical)
metric and $m_{\rm g}$ is the graviton mass.

The weak field limit is obtained with
$g_{\alpha\beta}=(g_0)_{\alpha\beta}+h_{\alpha\beta}$ and
$|h_{\alpha\beta}|\ll |(g_0)_{\alpha\beta}|$. In reality, as
discussed by \cite{matt98}, any algebraic function of the physical
metric and background metric with correct linearized behaviour up
to second order in $h$ would present the same characteristics of
equation (\ref{massa}).

Then, in the weak field limit, $T_{\mu\nu}^{\rm mass}$ becomes

\begin{equation}
T_{\mu\nu}^{\rm mass}=-\frac{m_{\rm g}^{2}c^{6}}{8\pi
G\hbar^{2}}\biggl\{h_{\mu\nu}-\frac{1}{2}\biggl[(g_0^{-1})^{\alpha\beta}
h_{\alpha\beta}\biggr](g_0)_{\mu\nu}\biggr\}.
\label{massivetensor}
\end{equation}

The field equations can be rearranged to produce a structure like
the usual Einstein field equations. That is,

\begin{equation}
R^{\mu\nu}-\frac{1}{2}g^{\mu\nu}R=-\frac{8\pi
G}{c^{4}}T^{\mu\nu}-\frac{8\pi G}{c^{4}}T^{\mu\nu}_{\rm
mass},\label{field}
\end{equation}

\noindent where $T^{\mu\nu}$ represents the usual energy-momentum
tensor.

From the last equation it is possible to verify that when $m_{\rm
g}\rightarrow 0$, we recover the ordinary Einstein field
equations.

\section{Weak gravitational waves with a massive graviton}
\label{sec3}

Considering gravitational waves far from field sources, we can
work in the weak-field limit. In this case, we should have to
choose a particular background geometry for the non-dynamical
metric. The natural choice, in a first work, is to take
$(g_0)_{\mu\nu}$ to correspond to a flat spacetime (Minkowski
spacetime). In principle, this choice should produce a good
agreement with all astrophysical observations at the level of weak
gravitational fields. Thus,

\begin{equation}
g_{\mu\nu}=\eta_{\mu\nu}+h_{\mu\nu},
\end{equation}

\noindent where $|h_{\mu\nu}|\ll 1$, $\eta_{\mu\nu}$ is the
Minkowski spacetime metric and the metric signature is
$(-1,+1,+1,+1)$.

Far from gravitational sources, we can take $T^{\mu\nu}=0$ (vacuum
solution) and we have from the conservation of stress-energy

\begin{equation}
\nabla_{\mu}G^{\mu\nu}=-\frac{8\pi
G}{c^{4}}\,\nabla_{\mu}T^{\mu\nu}_{\rm mass},
\label{conservation1}
\end{equation}

\noindent where $G^{\mu\nu}$ is the Einstein tensor and $\nabla$
denotes the covariant derivative.

Thus, we have for the right-hand side of equation
(\ref{conservation1})

\begin{equation}
{m_{\rm g}^2\, c^2\over \hbar^2}\,
\nabla_{\nu}\biggl(h^{\mu\nu}-\frac{1}{2}\eta^{\mu\nu}h\biggr)=0,
\label{conservation2}
\end{equation}

\noindent where we used in the last equation the result obtained
from equation (\ref{massivetensor}).

It is worth stressing that equation (\ref{conservation2}) is not a
gauge choice but a constraint imposed by the conservation of
energy.

On the other hand, the Ricci tensor is to first order in $h$

\begin{equation}
R_{\mu\nu}\simeq {\partial\over
\partial\,x^{\nu}}\Gamma^{\alpha}_{\alpha\mu}-{\partial\over
\partial\,x^{\alpha}}\Gamma^{\alpha}_{\mu\nu}+{\rm O}(h^{2}),
\end{equation}

\noindent and the affine connection is

\begin{equation}
\Gamma^{\alpha}_{\mu\nu}={1\over
2}\eta^{\alpha\beta}\biggl[{\partial\over
\partial\,x^{\mu}}h_{\beta\nu}+{\partial\over \partial\,x^{\nu}}h_{\beta\mu}
-{\partial\over \partial\,x^{\beta}}h_{\mu\nu}\biggr]+{\rm
O}(h^{2}).
\end{equation}

Thus, equation (\ref{field}) can be rewritten as

\begin{equation}
\square\biggl(h_{\mu\nu}-\frac{1}{2}\eta_{\mu\nu}h\biggr)
-m^{2}\biggl(h_{\mu\nu}-\frac{1}{2}\eta_{\mu\nu}h\biggr)=0,
\label{wave1}
\end{equation}

\noindent where we defined $m^2=m_{\rm g}^{2}\,c^{2}/{\hbar^{2}}$.

The last equation rewritten in terms of the trace reverse of
$h_{\mu\nu}$ produces

\begin{equation}
\square\,\bar{h}_{\mu\nu}-m^{2}\bar{h}_{\mu\nu}=0, \label{wave2}
\end{equation}

\noindent where $\bar{h}_{\mu\nu}=h_{\mu\nu}-1/2\eta_{\mu\nu}h$.

The general solution of equation (\ref{wave2}) is a linear
superposition of solutions of the form

\begin{equation}
\bar{h}_{\mu\nu}=e_{\mu\nu}\exp(i\,k_{\alpha}\,x^{\alpha}),
\label{wavesolution}
\end{equation}

\noindent where $e_{\mu\nu}$ is the polarization tensor.

From the normalization condition $k_{\mu}\,k^{\mu}=-m^{2}$, we can
obtain the dispersion relation

\begin{equation}
k=\sqrt{(\omega / c)^{2}-m^{2}}.
\end{equation}

As a consequence of massive gravitons, the speed of propagation of
a gravitational wave is dependent on the frequency and given by
$v(\omega)=d\omega/dk$ which produces

\begin{equation}
v(\omega)=c\sqrt{1-{m^2\,c^2\over \omega^2}}.
\end{equation}

It is important to stress that we are considering in our study a
gravitational wave travelling in the $+z$-direction. Thus, the
wave vector is given by $k_{\rm z}=k$ and the metric perturbation
presented in equation (14) may be rewritten as

\begin{equation}
\bar{h}_{\mu\nu}=e_{\mu\nu}\exp(-i\,\omega\,t+i\,k\,z).
\end{equation}

In the next section we will introduce the tetrad formalism to
determine the polarization wave modes in this massive theory.

\section{Polarization states of gravitational waves with a massive graviton}
\label{sec4}

For nearly null gravitational waves in the weak field limit, it is
only necessary to restrict our study to the form and behaviour of
the Riemann tensor. It is the Riemann tensor that gives us the
information how a gravitational wave interacts with a detector.

To analyse the components of the Riemann tensor into independent
wave modes in as invariant a manner as possible, we should
investigate the transformation properties of the Riemann tensor
under Lorentz transformations which leave the wave direction
fixed.

Then, we choose a basis vectors in which the components of the
Riemann tensor are computed. The basis vectors form a
quasiorthonormal tetrad basis. In particular, we follow the
formalism derived by \cite{newmannpenrose} and used in the work of
\cite{eardleylightmanlee,lightmanlee}.

It is worth stressing that the original tetrad formalism differs
from the Newman$-$Penrose formalism only in the manner of choice
of the basis vectors. That is, instead of an orthonormal basis,
the choice is made of a complex null-basis $(k,l,m,\bar{m})$ where
$k$ and $l$ are two real null-vectors and $m$ and $\bar{m}$ are a
pair of complex conjugate null-vectors.

These vectors satisfy the orthogonality relations

\begin{equation}
k\cdot\,l=1,\;\;\;\; m\cdot\,\bar{m}=-1,\;\;\;\;
k\cdot\,m=k\cdot\,\bar{m}=l\cdot\,m=l\cdot\,\bar{m}=0.
\label{normalization}
\end{equation}

In particular, $k$ and $l$ are tangent to the propagation
directions of the two plane waves, a wave propagating in the $+z$
direction and a wave propagating in the $-z$ direction,
respectively.

Following \cite{eardleylightmanlee,lightmanlee} and taking into
account the normalization conditions in equation
(\ref{normalization}), the vectors $(k,l,m,\bar{m})$ can be
written as

\begin{equation}
k={-\frac{1}{\sqrt{2}}}(1,0,0,1),
\end{equation}

\begin{equation}
 l={-\frac{1}{\sqrt{2}}}(1,0,0,-1),
\end{equation}

\begin{equation}
m={-\frac{1}{\sqrt{2}}}(0,1,{\rm i}\,,0),
\end{equation}

\begin{equation}
\bar{m}={-\frac{1}{\sqrt{2}}}(0,1,-{\rm i}\,,0).
\end{equation}

Although this basis vector form a null-basis, it is possible to
expand the wave vector $k$ for a gravitational wave not exactly
null \cite{will93}, like that obtained from the bimetric theory
studied here.

In the Newman$-$Penrose formalism, the Riemann tensor is split
into irreducible parts: the Weyl tensor, the traceless Ricci
tensor and the Ricci scalar named, respectively, tetrad components
$\Psi$, $\Phi$, and $\Lambda$.

The ten independent components of the Weyl tensor are represented
by the five complex scalars,

\begin{equation}
\Psi_{0}=-C_{pqrs}k^{p}\,m^{q}\,k^{r}\,m^{s}\;,
\end{equation}

\begin{equation}
\Psi_{1}=-C_{pqrs}k^{p}\,l^{q}\,k^{r}\,m^{s}\;,
\end{equation}

\begin{equation}
\Psi_{2}=-C_{pqrs}k^{p}\,m^{q}\,\bar{m}^{r}\,l^{s}\;,
\end{equation}

\begin{equation}
\Psi_{3}=-C_{pqrs}k^{p}\,l^{q}\,\bar{m}^{r}\,l^{s}\;,
\end{equation}

\begin{equation}
\Psi_{4}=-C_{pqrs}l^{p}\,\bar{m}^{q}\,l^{r}\,\bar{m}^{s}\;.
\end{equation}

It is also possible to define the following scalars representing
the Ricci tensor

\begin{equation}
\Phi_{00}=-{1\over 2}\,R_{kk}\;,
\end{equation}

\begin{equation}
\Phi_{01}=\Phi^{\ast}_{10}=-{1\over 2}\,R_{km}\;,
\end{equation}

\begin{equation}
\Phi_{11}=-{1\over 4}\,R_{kl}+R_{m\bar{m}}\;,
\end{equation}

\begin{equation}
\Phi_{12}=\Phi^{\ast}_{21}=-{1\over 2}\,R_{lm}\;,
\end{equation}

\begin{equation}
\Phi_{22}=-{1\over 2}\,R_{ll}\;,
\end{equation}

\begin{equation}
\Phi_{02}=\Phi_{20}=-{1\over 2}\,R_{mm}\;,
\end{equation}

\noindent and

\begin{equation}
\Lambda={R\over 24}={(R_{kl}-R_{m\bar{m}})\over 12}.
\end{equation}

In GR where $v=c$ only the component $\Psi_{4}$ is not null.
However, in the bimetric theory studied here, we observe that for
gravitational waves with $v(\omega)\;{^<_\sim}\;c$, the tetrad
components are $\Psi_{0}\simeq {\rm
O}(\varepsilon^2\,R),\;\Psi_{1}\simeq {\rm
O}(\varepsilon\,R),\;\Phi_{00}\simeq {\rm
O}(\varepsilon^2\,R),\;\Phi_{01}\simeq
\Phi_{10}\simeq\Phi_{02}\simeq\Phi_{20}\simeq{\rm
O}(\varepsilon\,R),\;\Phi_{11}=3/2\,\Psi_{2},\;\Phi_{12}=\Psi^{\ast}_{3},$
where $\varepsilon$ is related to the difference in speed between
light and the propagating gravitational wave. That is,
$\varepsilon=(c/v(\omega))^2-1$.

Thus, to describe the six independent components of Riemann tensor
we shall choose the set $\Psi_{2},\;\Psi_{3},\;\Psi_{4}$ and
$\Phi_{22}$.

It is important to stress that for low frequency gravitational
waves where $\varepsilon >1$, the componentes
$\Psi_{0},\;\Psi_{1},\;\Phi_{00},\;\Phi_{01},\;
\Phi_{10},\;\Phi_{02},\;\Phi_{20}$ can be written in function of
$\Psi_{2},\;\Psi_{3},\;\Psi_{4}$ and $\Phi_{22}$. Thus, these six
independent components are able to describe completely the
polarization modes of a gravitational wave.

Then, the explicit expressions for these six components are

\begin{equation}
\Psi_{2}={1\over
12}{h_{33}(\omega^{4}-\omega^{2}\,k^{2})+h_{00}(k^{4}-\omega^{2}\,k^{2})\over
(\omega^{2}\,+k^{2})},
\label{psi2}
\end{equation}

\begin{equation}
\Psi_{3}={1\over
8}{h_{13}(\omega^{3}+\omega^{2}\,k-\omega\,k^{2}-k^{3})+{\rm
i}\,h_{23} (\omega^{3}+\omega^{2}\,k-\omega\,k^{2}-k^{3})\over
\omega}, \label{psi3}
\end{equation}

\begin{eqnarray}
\Psi_{4} &=& {1\over
8}[{h_{00}(-\omega^{4}-2\,\omega^{3}\,k+2\,\omega\,k^{3}+k^{4})} \
\nonumber\\ &&
-{h_{22}(2\,\omega^{4}+4\,\omega^{3}\,k+4\,\omega^{2}\,k^{2}+4\,\omega\,k^{3}+2\,k^{4})}
\ \nonumber\\ &&
+{h_{33}(-\omega^{4}-2\,\omega^{3}\,k+2\,\omega\,k^{3}+k^{4})} \
\nonumber\\ && + {{\rm i}\,h_{12} (2\,\omega^{4}+4\,\omega^{3}\,k
+4\,\omega^{2}\,k^{2}+4\,\omega\,k^{3}
+2\,k^{4})}]/(\omega^{2}+k^{2}), \label{psi4}
\end{eqnarray}

\bigskip

\begin{equation}
\Phi_{22} = {1\over
8}{(h_{00}+h_{33})(-\omega^{4}-2\,\omega^{3}\,k+2\,\omega\,k^{3}+k^{4})\over
(\omega^{2}+k^{2})},
\label{phi22}
\end{equation}

\noindent where we use in the above set of equations $c=1$.

In figure 1 we show the displacement that each mode induces on a
sphere of test particles. The wave propagates in the $+z$
direction and has time dependence $\cos\,\omega\,t$. The solid
line is an instantaneous at $\omega\,t=0$ and the broken line is
one at $\omega\,t=\pi$.

\begin{figure}
\begin{center}
\leavevmode
\centerline{\epsfig{figure=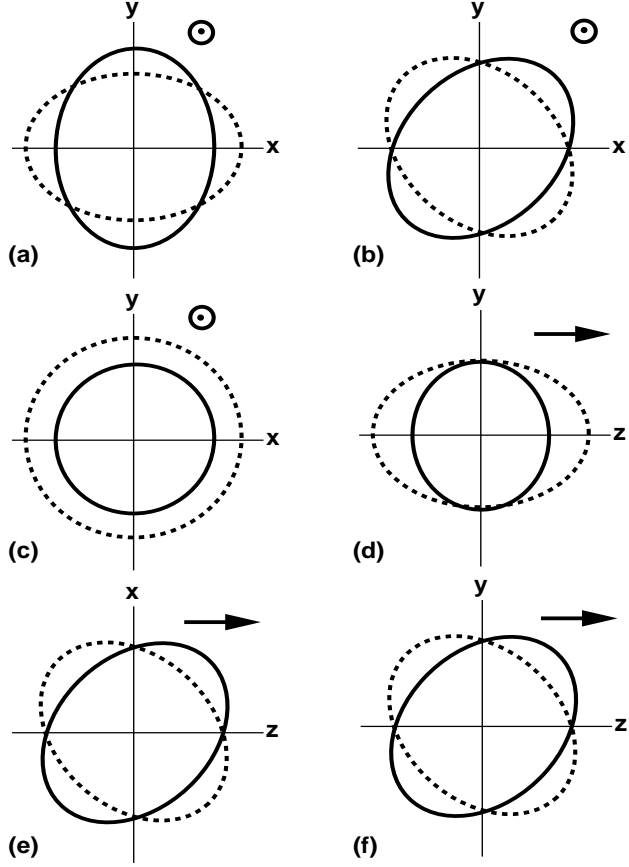,angle=360,height=11.5cm,width=8.25cm}}
\caption{The six polarization modes of weak gravitational waves
permitted in any theory of gravity. Shown is the displacement that
each mode induces on a sphere of test particles. The wave
propagates out of the plane in (a), (b), (c), and it propagates in
the plane in (d), (e) and (f). The displacement induced on the
sphere of test particles corresponds to the following
Newman$-$Penrose quantities: ${\rm Re}\;\Psi_{4}$ (a), ${\rm
Im}\;\Psi_{4}$ (b), $\Phi_{22}$ (c), $\Psi_{2}$ (d), ${\rm
Re}\;\Psi_{3}$ (e), ${\rm Im}\;\Psi_{3}$ (f).}
\end{center}
\label{fig1}
\end{figure}

We can solve the set of equations (\ref{psi2}), (\ref{psi3}),
(\ref{psi4}), (\ref{phi22}) for determined gravitational wave
frequencies, in order to identify the contribution of the extra
polarization modes for a signal received by a gravitational wave
detector.

In particular, we present the results for three different
frequencies: $f_{\rm GW}=100\,{\rm Hz}$ that corresponds
approximately to the frequency of maximum sensibility for the LIGO
interferometer; $f_{\rm GW}=10^{-3}\,{\rm Hz}$ that corresponds to
the maximum sensibility for the future laser interferometer space
antenna (LISA) and $f_{\rm GW}\simeq 10^{-7}\,{\rm Hz}$.

Thus, we have $m_{\rm g}= 0.44 \times 10^{-21}{\rm eV}/c^2$ that
corresponds to the best limit from solar system observations
\cite{talmadge}.

a) $f_{\rm GW}=100\,{\rm Hz}$:

\begin{equation}
\Psi_{2}\simeq 2.1\times 10^{-35}(h_{33}-h_{00}), \label{psi2f100}
\end{equation}

\begin{equation}
\Psi_{3}\simeq 1.2\times 10^{-34}(h_{13}+{\rm i}\,h_{23}),
\label{psi3f100}
\end{equation}

\begin{equation}
\Psi_{4}\simeq 1.2\times 10^{-34}(-h_{00}-h_{33})+4.4\times
10^{-16}(-h_{22}+{\rm i}\,h_{12}), \label{psi4f100}
\end{equation}

\begin{equation}
\Phi_{22}\simeq 1.2\times 10^{-34}(-h_{00}-h_{33}).
\label{phi22f100}
\end{equation}

b) $f_{\rm GW}=10^{-3}\,{\rm Hz}$:

\begin{equation}
\Psi_{2}\simeq 2.1\times 10^{-35}(h_{33}-h_{00}),
\label{psi2flisa}
\end{equation}

\begin{equation}
\Psi_{3}\simeq 1.2\times 10^{-34}(h_{13}+{\rm i}\,h_{23}),
\label{psi3flisa}
\end{equation}

\begin{equation}
\Psi_{4}\simeq 1.2\times 10^{-34}(-h_{00}-h_{33})+4.4\times
10^{-26}(-h_{22}+{\rm i}\,h_{12}), \label{psi4flisa}
\end{equation}

\begin{equation}
\Phi_{22}\simeq 1.2\times 10^{-34}(-h_{00}-h_{33}).
\label{phi22flisa}
\end{equation}

c) $f_{\rm GW}=1.1\times 10^{-7}\,{\rm Hz}$:

\begin{equation}
\Psi_{2}\simeq -2.1\times 10^{-36}h_{00}+3.9\times 10^{-35}h_{33},
\label{psi2fmg}
\end{equation}

\begin{equation}
\Psi_{3}\simeq 7.6\times 10^{-35}(h_{13}+{\rm i}\,h_{23}),
\label{psi3fmg}
\end{equation}

\begin{equation}
\Psi_{4}\simeq 8.9\times 10^{-35}(-h_{00}-h_{33})+2.0\times
10^{-34}(-h_{22}+{\rm i}\,h_{12}), \label{psi4fmg}
\end{equation}

\begin{equation}
\Phi_{22}\simeq 8.9\times 10^{-35}(-h_{00}-h_{33}).
\label{phi22fmg}
\end{equation}

\section{Final remarks}
\label{sec5}

Using the bimetric theory proposed by \cite{matt98}, we obtain the
explicit expressions for the polarization modes of gravitational
waves considering a nonzero mass for the graviton. In particular,
we analyse what happens to the six tetrad independent components
for three different frequencies, namely, $f_{\rm GW}=100\,{\rm
Hz}$ approximately the frequency of maximum sensibility for the
LIGO interferometer; $f_{\rm GW}=10^{-3}\,{\rm Hz}$ the frequency
of maximum sensibility for the future laser interferometer space
antenna (LISA) and $f_{\rm GW}\simeq 10^{-7}\,{\rm Hz}$.

From the above results, we can see that for $f_{\rm GW}=100\,{\rm
Hz}$ the dominant tetrad component is $\Psi_{4}$, with the part
purely GR several orders of magnitude bigger than the extra
polarization terms. For example, $\Psi_{4}$ can be split into
$\Psi_{4}=(\Psi_{4})_{\rm{GR}}+(\Psi_{4})_{\rm{M}}$, where the
polarizations purely GR are given by $(\Psi_{4})_{\rm{GR}}\simeq
4.4\times 10^{-16}(-h_{22}+{\rm i}\,h_{12})$ while, the term due
to the nonzero mass for the graviton is $(\Psi_{4})_{\rm{M}}\simeq
1.2\times 10^{-34}(-h_{00}-h_{33})$.

If we consider that all metric perturbations $h_{\mu\nu}$ have a
similar value $h$ then we obtain $|(\Psi_{4})_{\rm{GR}}|\simeq
10^{18}\,|(\Psi_{4})_{\rm{M}}|$. Thus, the term
$(\Psi_{4})_{\rm{M}}$ has behaviour similar to a very small
perturbation when compared to the GR term in $\Psi_{4}$. A similar
conclusion can be obtained from the comparison of the amplitudes
of gravitational waves in equations (\ref{psi2f100}),
(\ref{psi3f100}), (\ref{psi4f100}) and (\ref{phi22f100}). We can
see that the modes $\Psi_{2}$, $\Psi_{3}$ and $\Phi_{22}$ maintain
the same very small perturbation behaviour as the component
$(\Psi_{4})_{\rm{M}}$.

Then, at least for $m_{\rm g}= 0.44 \times 10^{-21}{\rm eV}/c^2$
and taking into account the weak field limit studied in the
present paper, it should not be possible to identify the signature
of the extra polarization modes from an astrophysical source
detected, for example, by VIRGO and LIGO interferometers.
Certainly, as mentioned above, this conclusion is based on the
fact that all metric perturbations $h_{\mu\nu}$ have a similar
value $h$. In this case, the extra polarization modes are very
small perturbations of $(\Psi_{4})_{\rm GR}$.

For the case $f_{\rm GW}=10^{-3}\,{\rm Hz}$, we can see that the
extra polarization modes have the same amplitudes as in the
frequency $f_{\rm GW}=100\,{\rm Hz}$. However, the weight of the
term purely GR in $\Psi_{4}$ decreases when compared to the
massive term $(\Psi_{4})_{\rm{M}}$. For example, from equation
(\ref{psi4flisa}) we obtain that $|(\Psi_{4})_{\rm{GR}}|\simeq
10^{8}\,|(\Psi_{4})_{\rm{M}}|$ for $f_{\rm GW}=10^{-3}\,{\rm Hz}$.

As a consequence, we could think that, in principle, a
gravitational wave signal in the LISA frequency range could give
us more information on the polarization modes $\Psi_{2}$,
$\Psi_{3}$ and $\Phi_{22}$ than a gravitational wave signal in the
VIRGO-LIGO frequency range. But unfortunately, same as for $f_{\rm
GW}=10^{-3}\,{\rm Hz}$ the extra polarization modes correspond to
very small perturbations of the GR polarizations.

On the other hand, a very interesting result appears for  the
frequency $f_{\rm GW}=1.1\times 10^{-7}\,{\rm Hz}$. In this case,
we can see from equations (\ref{psi2fmg}), (\ref{psi3fmg}),
(\ref{psi4fmg}) and (\ref{phi22fmg}) that all tetrad components
have similar amplitude. As a consequence, we have
$|(\Psi_{4})_{\rm{GR}}|\simeq |(\Psi_{4})_{\rm{M}}|$. In
particular, all the polarization modes produce similar
`excitation' in the frequency $1.1\times 10^{-7}\,{\rm Hz}$.

This kind of result is maintained if we change the value of the
graviton mass. However, if we reduce the value of the graviton
mass, the main result is to shift the frequency $f_{\rm c}$ where
all the polarization modes have similar amplitude. That is, lower
the graviton mass, lower the frequency $f_{\rm c}$ is.

In figure 2 we can see the relation between the mass of the
graviton and the frequency $f_{\rm c}$. For example, if the mass
of the graviton is $m_{\rm g}= 0.44 \times 10^{-21}{\rm eV}/c^2$,
then we have $f_{\rm c}\simeq 10^{-7}\,{\rm Hz}$. On the other
hand, if the mass of the graviton is $m_{\rm g}= 0.44 \times
10^{-29}{\rm eV}/c^2$ then $f_{\rm c}\simeq 10^{-15}\,{\rm Hz}$.

\begin{figure}
\begin{center}
\leavevmode
\centerline{\epsfig{figure=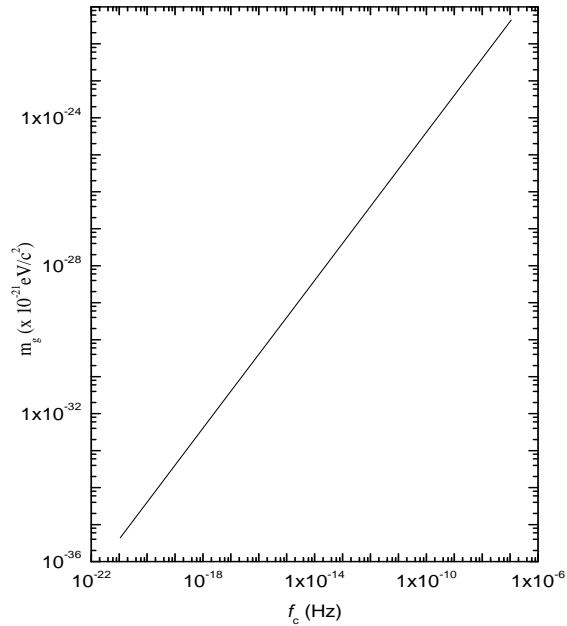,angle=360,height=10.25cm,width=8.25cm}}
\caption{The relation between the mass of the graviton and the
frequency where all the polarization modes presented in figure 1
have the same amplitude.}
\end{center}
\label{fig2}
\end{figure}

This result could be used to obtain a new limit on the graviton
mass based on the analysis of the maps produced by experiments
that study the anisotropy of the cosmic microwave background
(CMB).

In particular, the generation of a stochastic background of
primordial gravitational waves is a fundamental prediction of
inflationary models for the early universe. Its amplitude is
determined by the energy scale of inflation, which can widely vary
between different inflationary models.

Gravitational wave detectors, however, are quite unlikely to have
enough sensitivity to detect such a primordial signal, owing both
to its smallness and to its extremely low characteristic
frequencies. The existence of ultra-low-frequency gravitational
radiation, however, can be indirectly probed thanks to the
temperature anisotropy and polarization it induces on the CMB
radiation.

In particular, the curl component, called B-mode, of the CMB
polarization provides a unique opportunity to disentangle the
effect of tensor (gravitational-wave) from scalar perturbations,
as this is only excited by either tensor or vector modes
\cite{seljak,kamionkowski}.

From this point of view, future satellite missions, such as
Planck, which will have enough sensitivity to either detect or
constrain the B-mode CMB polarization predicted by the simplest
inflationary models, might represent the first 'space-based
gravitational-wave detector' \cite{caldwell}.

Thus, future CMB missions could present an alternative way to
impose a new upper limit on the mass of the graviton and to
constrain the number of polarization modes of the gravitational
waves.

\ack{We thank Drs O D Aguiar and J C N de Araujo for stimulating
discussions. WLSP and ODM would like to thank the Brazilian agency
FAPESP for support (grants 02/13423-2; 02/07310-0 and 02/01528-4,
respectively). RMM also thanks the Brazilian agency FAPESP for
financial support (grant 99/10809-2).}

\end{document}